\documentclass{article}

\usepackage{arxiv}
\usepackage{hyperref}
\usepackage{graphics}
\usepackage{graphicx}
\usepackage{dcolumn}
\usepackage{bm}
\usepackage{amsmath}
\usepackage{amssymb}
\usepackage{xcolor}
\usepackage{dsfont}
\usepackage{braket}
\numberwithin{equation}{section}

\begin{document}

\title{A neutrino data analysis of extra-dimensional theories with massive bulk fields}

\author{Philipp Eller \\
\href{mailto:philipp.eller@tum.de}{philipp.eller@tum.de} \\
Technical University Munich (TUM)\\
James-Franck-Strasse 1\\
85748 Garching, Germany
\And
Manuel Ettengruber\\
\href{mailto:manuel-meinrad.ettengruber@cea.fr}{manuel-meinrad.ettengruber@cea.fr}\\
Universite Paris-Saclay,\\
CNRS, CEA, Institut de Physique Theorique\\
91191, Gif-sur-Yvette, France
\And
Alan Zander\\
\href{mailto:alan.zander@tum.de}{alan.zander@tum.de}\\
Technical University Munich (TUM)\\
James-Franck-Strasse 1\\
85748 Garching, Germany}

\maketitle

\begin{abstract}
We present a global neutrino oscillation analysis of models with a single large extra dimension in which right‑handed neutrinos possess bulk Dirac masses. Two scenarios are considered: Large Extra Dimensions with bulk masses and the Dark Dimension framework, both predicting a tower of sterile Kaluza–Klein states that mix with active neutrinos. Using data from MINOS/MINOS+, KamLAND, and Daya Bay, we perform a joint likelihood analysis. No signatures of these theories were found. Therefore, we constrain the compactification radius under different bulk mass and Yukawa coupling assumptions. Large positive bulk masses or sizable Yukawas lead to strong bounds, while small couplings or negative bulk masses remain less constrained.
\end{abstract}

\section{INTRODUCTION \label{sec: Introduction}} 
The idea that our universe may contain extra spatial dimensions beyond the familiar three has a long and compelling history. Originally introduced in attempts to unify gravity with other fundamental forces, extra dimensions have since become central also to a number of further influential theoretical frameworks, most notably string theory. These models are motivated by diverse goals: resolving the hierarchy problem \cite{Arkani-Hamed:1998jmv, Randall:1999ee}, explaining the nature of dark matter \cite{Arkani-Hamed:1998sfv, Gonzalo:2022jac, Friedlander:2022ttk,Anchordoqui:2022txe, Ettengruber:2025kzw} and dark energy \cite{Montero:2022prj}, constructing consistent theories of quantum gravity \cite{Scherk:1974ca, Antoniadis:1998ig}, and generating small neutrino masses naturally \cite{Arkani-Hamed:1998wuz, Anchordoqui:2022svl}. Numerous studies have explored the phenomenological implications of extra dimensions, often revealing novel signatures in particle physics and cosmology.


Models with Large Extra Dimensions (LED) were originally proposed to resolve the electroweak hierarchy problem by lowering the fundamental scale of gravity $M_f$ to the TeV range \cite{Arkani-Hamed:1998jmv}. This happens due to the large number of resulting Kaluza-Klein (KK) modes coming from the compactification of spatial dimensions which separates the fundamental scale of gravity from the Planck scale \cite{Dvali:2007hz, Dvali:2007wp}. In these scenarios, hypothetical right-handed neutrinos are allowed to propagate in the bulk, which leads to small active neutrino masses due to volume suppression \cite{Arkani-Hamed:1998wuz, Dvali:1999cn} which is nowadays understood as a specific realization of the many mixing partners mechanism \cite{Ettengruber:2022pxf, Ettengruber:2025usk}. These models have been extended to include bulk Dirac mass terms, giving rise to what we refer to as ADD+ (sometimes also referred as LED+ in the literature) scenarios \cite{Lukas:2000wn, Agashe:2000rw, Diego:2008zu}. Such bulk masses may help explain anomalies observed in short-baseline experiments like LSND and MiniBooNE \cite{Carena:2017qhd}.

More recently, the persistent smallness of the observed cosmological constant $\Lambda$ and developments in the Swampland program have pointed toward the possibility of a mesoscopic extra dimension, with a length scale on the order of microns. This idea, known as the "dark dimension" (DD) \cite{Montero:2022prj}, emerges from quantum gravity conjectures combined with observational data, especially the dark energy scale \cite{Vafa:2005ui, Ooguri:2006in, Lee:2019xtm, Lust:2019zwm, Lee:2019wij}. In this framework, the effective field theory of our universe is expected to break down at a fundamental scale of gravity, or species scale, around $M_f \sim 10^9$ GeV. A tower of light states, including KK gravitons and sterile (right-handed) neutrinos, becomes relevant at an energy scale close to the dark energy scale, $\Lambda^{1/4} \sim$ meV. Relating the species scale $M_f$ to the dark energy scale leads to a concrete prediction for the compactification radius $R$ of the largest extra dimension, which characterizes the mesoscopic extra-dimensional length scale. 
In the dark dimension framework, the compactification radius is expected to lie in the range $R \sim 0.1 - 10 \mu$m. In the standard case without bulk neutrino masses, experimental limits require $R \lesssim 0.2 \mu$m \cite{forero2022large, machado2011testing}, exerting great pressure on the dark dimension. Allowing for non‑zero bulk Dirac masses relaxes these bounds, substantially expanding the viable parameter space.

The introduction of bulk masses for sterile neutrinos serves multiple purposes. From a phenomenological standpoint, they allow for greater flexibility in shaping the KK mass spectrum and the mixing pattern with active neutrinos. In particular, bulk masses can suppress the coupling to the lowest KK modes, thereby evading strong constraints from disappearance experiments while still permitting observable appearance signals. This feature has been shown to potentially reconcile the tension between appearance anomalies and null disappearance results \cite{Carena:2017qhd}. Additionally, bulk masses can naturally arise in a variety of higher-dimensional setups, where their structure is tied to the geometry or dynamics of the extra-dimensional space \cite{Anchordoqui:2023wkm}.

Despite their different theoretical motivations, both ADD+ and DD scenarios share a common structure: sterile neutrinos propagating in a flat extra dimension, with non-zero bulk masses and brane-localized Yukawa couplings. This leads to a tower of sterile KK modes that mix with active neutrinos and can influence oscillation probabilities at eV-scale mass splittings. The main difference lies in the value and interpretation of the fundamental scale of gravity $M_f$: ADD+ imposes $M_f \sim $TeV, while being independent of $R$ by invoking the existence of additional smaller extra dimensions that, overall, substantially contribute to the extra-dimensional volume. On the other hand, in the DD model, $M_f$ has a dependency on $R$ and is related to the size of the cosmological constant, resulting in $M_f \sim 10^9$ GeV.

In this work, we present an analysis that applies to both ADD+ and DD scenarios, emphasizing their shared theoretical and statistical framework while carefully distinguishing their interpretations of the fundamental scale of gravity. We study the resulting mixing patterns and derive constraints on the compactification radius $R$ and bulk mass parameters using existing neutrino oscillation data.

The rest of the paper is organized as follows. In Section II, we introduce the formalism of neutrino propagation in extra dimensions, allowing for bulk Dirac mass terms for the right-handed neutrinos. In Section III, we discuss the phenomenological consequences of the resulting Kaluza-Klein spectrum and analyze its impact on neutrino oscillation probabilities. In Section IV, we explain how we conducted the analysis, while in Section V we show the obtained results. Section VI concludes with a discussion of the implications for both the ADD+ and DD scenarios.

\section{MODEL\label{sec: model}}
We consider a model where three generations of right-handed neutrinos propagate in a flat five-dimensional spacetime, with the fifth dimension compactified on an $S^1/\mathbb{Z}_2$ orbifold of radius $R$, while the Standard Model (SM) fields remain confined to a four-dimensional brane.


This framework is common to both ADD+ and the DD scenarios. In both cases, we assume a single large extra dimension, where sterile neutrinos propagate with possible non-zero bulk Dirac masses.

\subsection{Fundamental scale of gravity}

One key feature of LED models is the strong separation between the fundamental scale of gravity $M_f$ and the Planck scale $M_{\text{Pl}}$. In \cite{Dvali:2007hz, Dvali:2007wp} it was shown that the presence of additional particle species in the spectrum causes this separation. In compactified extradimensional theories, a large number of KK-modes automatically arises, which then leads to this effect. For such extradimensional theories the separation of scales is expressed by  \cite{Arkani-Hamed:1998jmv, Dvali:1999cn}

\begin{equation}
    M_f = \left(\frac{M_P^2}{(2\pi)^n R_1 R_2 ... R_n}\right)^{\frac{1}{n+2}} \;,
    \label{masterequation}
\end{equation}
with $n$ being the number of extra dimensions and $R_1, R_2,...,R_n$ are the radii of these dimensions. From this equation, one can see in what sense ADD and DD scenarios are different. In ADD, the number of extra dimensions is $n \geq 2$, which means by introducing a hierarchy among the radii, one can create a phenomenology where the largest extra dimension plays a dominant role and one could treat it like a 5D scenario with one extra dimension. Nevertheless, because the other extra dimensions are still present, one can keep $M_f$ on a scale that is not fully determined by the size of the largest extra dimension. 

This is different for DD models. In this theory it is argued that there should be essentially just one extra dimension that is significantly larger than the Planck scale (for a variation with two relevant extra dimensions see \cite{Anchordoqui:2025nmb}). Therefore, due to \eqref{masterequation} the size of the extra dimension and $M_f$ is related. For one extra dimension, one can then bring it down to a more phenomenological user friendly form
\begin{equation}
M_f = 1.055 \cdot 10^9 \text{GeV} \left(\frac{\mu\text{m}}{2 \pi R}\right)^{1/3} \left(\frac{M_{\text{Pl}}}{2.435\cdot 10^{18} \text{GeV}}\right)^{2/3} \;,
\label{eq: Mf for DD}
\end{equation}
with $M_{\text{Pl}}$ being the reduced Planck mass.

\subsection{Fermion structure in the bulk}

The 5D action for a single generation of a SM-singlet fermion $\Psi_i$ includes kinetic terms, a bulk Dirac mass term, and Yukawa interactions. Under the assumption of minimal flavor violation, one can work in a basis of the 5D fermions $\Psi_i$ in which both the bulk Dirac mass terms and the 5D Yukawa couplings are diagonal, 

\begin{align} 
S_{\text{MFV}} & = - \int d^4x \, dz \sum_{i=1}^3 \bigg[c_i \, \overline{\Psi}_i \Psi_i + \notag \\
& \delta(z) \left( y_i \, \overline{L}_i(x) \tilde{H}(x) \Psi_i^R(x,z) + \text{h.c.} \right)
\bigg],
\end{align}

where $\tilde{H} = i \sigma_2 H^*$ is the charge-conjugated Higgs doublet, $L_i$ is the left-handed lepton doublet, and the $\delta$-function enforces that the Yukawa interactions are localized on the brane at $z = 0$.

Furthermore, since the fifth dimension is compact, the 5D field can be expressed as a Fourier series in terms of KK eigenmodes:
\begin{equation}
  \Psi_i(x,z) = \sum_{n=0}^\infty \left[ \psi_{i,n}^L(x) f_{i,n}^L(z) + \psi_{i,n}^R(x) f_{i,n}^R(z) \right],
\end{equation}
where $\psi_{i,n}^L$ and $\psi_{i,n}^R$ are the left- and right-handed 4D spinors, and $f_{i,n}^L(z)$, $f_{i,n}^R(z)$ are their corresponding wavefunctions along the compact fifth dimension. This KK decomposition allows the 5D fermion to be described as an infinite tower of 4D spinors with increasing masses. Each mode corresponds to a distinct sterile neutrino in the 4D effective theory, with masses determined by the compactification scale and the bulk Dirac mass.

The functions $f_{i,n}^L(z)$ and $f_{i,n}^R(z)$ are determined by solving the 5D Dirac equation with appropriate boundary conditions, and form a complete orthonormal basis over the compact interval. Imposing Dirichlet boundary conditions, we have

\begin{equation}
    f_{i, 0}^R (z) = \sqrt{\frac{2 c_i}{e^{2 \pi R c_i} - 1}} e^{c_i z},
\end{equation}

\begin{equation}
    f_{i, n}^L (z) = \sqrt{\frac{2}{\pi R}} \sin\left(\frac{n z}{R}\right),
\end{equation}

\begin{equation}
    f_{i, n}^R (z) = \sqrt{\frac{2}{\pi R m_{i, n}^2}} \left[c_i \sin\left(\frac{n z}{R}\right) + \frac{n}{R} \cos \left(\frac{n z}{R}\right)\right],
\end{equation}
with masses given by
\begin{equation}
m_{i, n} = \sqrt{\left(\frac{n}{R}\right)^2 + c_i^2},
\label{eq: KK masses + c}
\end{equation}

where $n = 1, 2, \dots$ labels the KK level, and $c_i$ is the bulk Dirac mass for generation $i$. Note that the masses from equation (\ref{eq: KK masses + c}) would correspond to the physical masses of the KK tower in the absence of interactions with the SM.

However, the brane-localized Yukawa couplings connect the SM lepton doublets to the KK tower. Their effective value can be derived by integrating over the extra dimension. Assuming again that the SM is localized at $z = 0$, the couplings are:
\begin{align}
\label{eq: Y0i} Y_0^i &= \lambda^i \frac{M_f}{M_{\text{Pl}}} \sqrt{\frac{2 \pi c_i R}{e^{2 \pi c_i R} - 1}}, \\
\label{eq: Yni} Y_n^i &= \lambda^i \frac{M_f}{M_{\text{Pl}}} \sqrt{\frac{2 n^2}{n^2 + c_i^2 R^2}},
\end{align}
where $\lambda^i$ are dimensionless Yukawa couplings.

N.B.: The sign of bulk mass terms $c_i$´s depends on the compactification scheme and the specific pick of boundary conditions of the bulk fields on the boundary \cite{Csaki:2005vy, Ponton:2012bi}. In our analysis, we explore both signs of the bulk mass term, as the sign determines whether the zero-mode is localized toward or away from the brane, leading to distinct physical scenarios.

\subsection{Mass matrix and mixing}

In our "intermediate basis", where the bulk mass and 5D Yukawa terms are diagonalized, the neutrino mass matrix for each generation $i$ takes the form
\begin{equation}
M_i = \begin{pmatrix}
v Y_0^i & 0      & 0      & \cdots & 0      \\
v Y_1^i & m_1^i  & 0      & \cdots & 0      \\
v Y_2^i & 0      & m_2^i  & \cdots & 0      \\
\vdots  & \vdots & \vdots & \ddots & \vdots \\
v Y_n^i & 0      & 0      & \cdots & m_n^i  \\
\end{pmatrix},
\end{equation}
where $v = 174$~GeV is the Higgs vacuum expectation value. The upper row corresponds to the brane-localized active neutrino, and the remaining rows to the sterile KK tower.

To obtain the full mass matrix relevant for neutrino oscillations, we must also rotate in flavor space using the usual PMNS matrix, which connects the weak interaction eigenstates to the light mass eigenstates. Thus, the complete mixing matrix is formed by diagonalizing the full mass matrix squared, while the physical squared masses are obtained from its eigenvalues.

\section{PHENOMENOLOGY \label{sec: pheno}}

The presence of sterile neutrinos propagating in a compact extra dimension leads to a rich phenomenology in neutrino oscillation experiments, as we already mentioned. In both the ADD+ and DD scenarios considered here, each 5D right-handed neutrino generates a tower of KK modes with masses and mixings governed by the compactification scale $R$ and the corresponding bulk mass parameters $c_i$. These towers couple to the SM neutrinos via brane-localized Yukawa interactions and lead to a number of distinct signatures that deviate from standard three-flavor oscillation predictions:

\begin{itemize}
\item {\bf Suppression of PMNS unitarity:} Mixing with sterile KK modes reduces the active flavor content of light neutrino states, leading to an effectively non-unitary PMNS matrix, creating deficits in disappearance channels.

\item {\bf New oscillation frequencies:} Each KK mode introduces a new mass-squared splitting $\Delta m^2_{n} \sim n^2/R^2 + c_i^2$, leading to additional oscillation frequencies.

\item {\bf Appearance signals:} In contrast to standard ADD (with $c_i = 0$), which primarily induces disappearance effects, models with nonzero $c_i$ allow for non-trivial active-to-active appearance channels, allowing $\nu_\mu \rightarrow \nu_e$ and other transitions at short baselines.
\end{itemize}

\begin{figure*}
 \centering
    \includegraphics[width=0.90\textwidth]{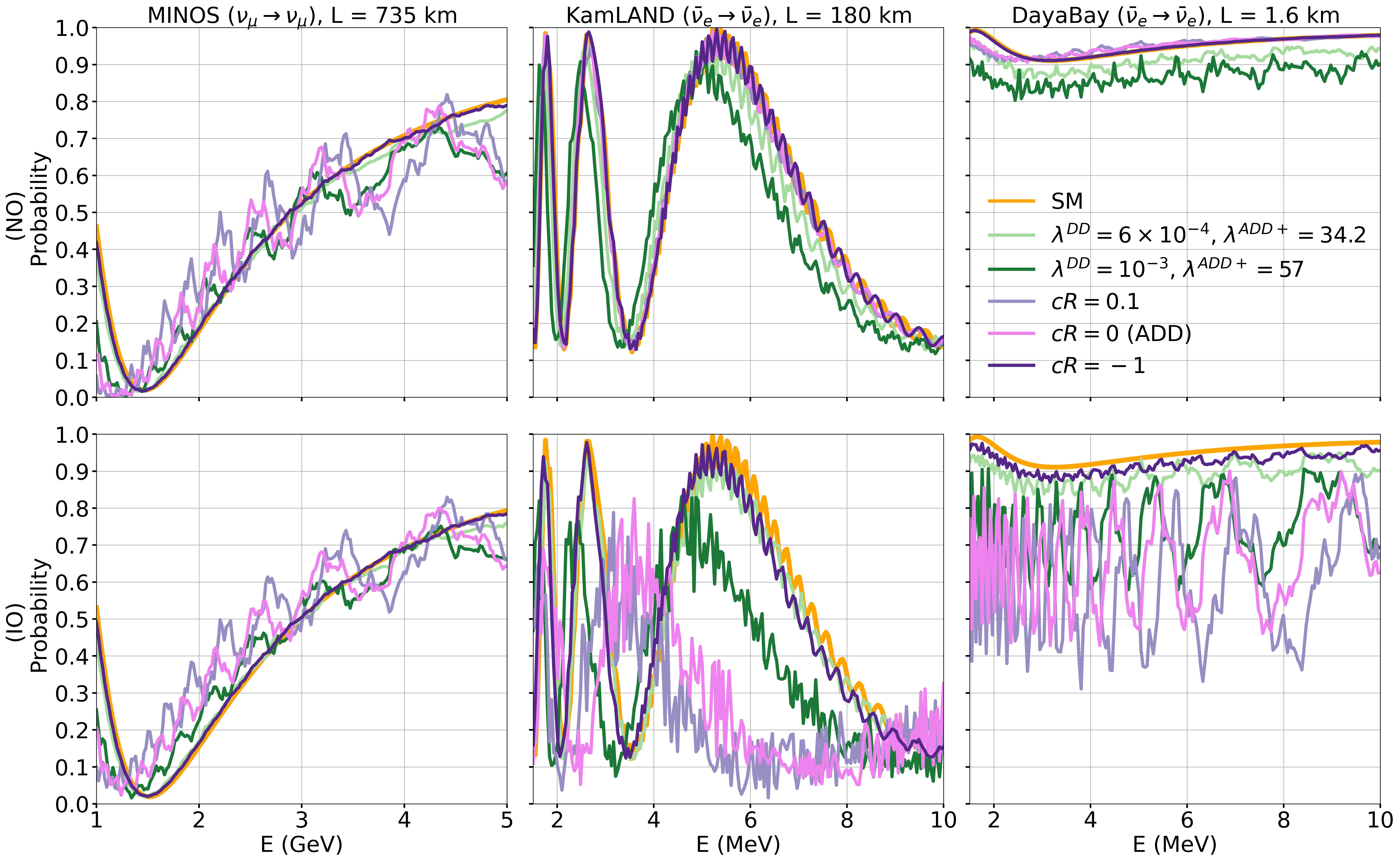}
    \caption{Oscillation probabilities for the three considered experiments, MINOS/MINOS+ (left), KamLAND (center), and Daya Bay (right), shown for both normal ordering (top row) and inverted ordering (bottom row). The curves compare the SM prediction (orange) to extra-dimensional scenarios with various parameter choices, where we have fixed the radius $R = 1 \mu$m, the lightest neutrino mass $m_{\text{lightest}} = 10^{-3}$ eV, and our computation was performed using $N_{KK} = 10$ KK modes. The $\lambda$-curves (green shades) correspond to different fundamental Yukawa couplings, while $c_iR$ were determined by the SM mass splittings, $\Delta m^2_{21}$ and $\Delta m^2_{31}$. For ADD+, we chose here $M_f = 10$ TeV. On the other hand, the $c_iR$-curves (purple shades) illustrate the effect of different bulk mass parameters, while choosing $\lambda_i$ to yield, again, the SM mass splittings. For the latter, there is no difference between ADD+ and DD.}
   \label{fig: oscillograms prob}
\end{figure*}

To validate our framework, we reproduced the oscillograms of Figure 4 from \cite{Carena:2017qhd}. Furthermore, we generated a series of representative oscillation probability curves for three key experimental setups that we will discuss in the next sections. These plots illustrate the energy-dependent survival probabilities across different model realizations and parameter choices, contrasting the predictions of the SM with those of the ADD+ and DD models. They can be found in Figure \ref{fig: oscillograms prob}, where we have fixed the radius at $R = 1 \mu$m, and the lightest neutrino mass at $m_{\text{lightest}} = 10^{-3}$ eV. To account for the difference in order of magnitude for the fundamental scale of gravity in both models, and still plot both models in one single diagram, we translate the dimensionless Yukawa coupling $\lambda$ from one model to the other using the relation $\lambda^{\text{ADD+}} M_f^{\text{ADD+}} = \lambda^{\text{DD}} M_f^{\text{DD}}$, where $M_f^{\text{DD}}$ is given in equation (\ref{eq: Mf for DD}) and $M_f^{\text{ADD+}}$ was chosen here to be $M_f^{\text{ADD+}} = 10$ TeV, for definiteness.


These signatures are complementary and allow multiple experimental avenues for probing the presence of large extra dimensions with sterile neutrinos. In particular, the eV-scale mass splittings naturally produced in this setup match the scales probed by current short- and long-baseline neutrino oscillation experiments.

\section{ANALYSIS \label{sec: analysis}}

In order to test the phenomenological viability of extra-dimensional models with bulk masses, we perform a global fit to neutrino oscillation data using a frequentist approach. In particular, we use a likelihood ratio test statistic to derive exclusion limits in the absence of significant deviations from the three-flavor paradigm.

Our analysis framework \footnote{https://github.com/philippeller/Newtrinos.jl} has been validated through its use in earlier studies \cite{ettengruber2024testing, kozynets2025constraints}, ensuring methodological soundness.
We include data from KamLAND, MINOS/MINOS+, and Daya Bay. These experiments are selected for their high statistics, complementary baselines and energy sensitivities, and well-understood systematic uncertainties. Together, they probe all oscillation sectors relevant to both the Standard Model and extra-dimensional scenarios:

\begin{itemize}
\item \textbf{MINOS/MINOS+}, with long baselines and broad energy coverage, is sensitive to $\nu_\mu$ disappearance and constrains $\Delta m^2_{31}$ and $\theta_{23}$;
\item \textbf{KamLAND}, a medium-baseline reactor experiment, offers excellent sensitivity to the solar parameters $\theta_{12}$ and $\Delta m^2_{21}$, which helps reduce degeneracies in global fits.
\item \textbf{Daya Bay}, with its short baseline and high-precision detectors, provides precise measurements of $\theta_{13}$ and $\Delta m^2_{31}$;
\end{itemize}

For MINOS and MINOS+, we use the combined dataset corresponding to $16.36 \cdot 10^{20}$ protons-on-target, in which muon neutrinos are generated at Fermilab and detected after traveling $L = 735$ km through the Earth to a far detector in northern Minnesota \cite{adamson2019search}. The KamLAND dataset includes $3.49 \cdot 10^{32}$ target-proton-years of exposure, detecting antineutrinos produced by dozens of nuclear reactors across Japan at an average baseline of $L \sim 180$ km using a kiloton-scale liquid scintillator detector located deep underground \cite{gando2011constraints}. From Daya Bay, we include the final legacy dataset comprising $3158$ days of operation and $5.55 \cdot 10^6$ inverse beta-decay candidates. Here, antineutrinos are generated by six reactor cores and measured at multiple detectors located 360–2000 meters away in underground halls near Shenzhen, China \cite{an2023precision}.

In our analysis, we perform a simultaneous profile likelihood fit including all relevant model parameters: the SM oscillation parameters $(\theta_{12}, \theta_{13}, \theta_{23}, \delta_{\rm CP}, \Delta m^2_{21}, \Delta m^2_{31}, m_{\text{lightest}})$, parameters involving systematic uncertainties, and the compactification radius $R$ of the extra dimension. To probe the structure of the bulk fermion sector, we consider two complementary approaches: in one, we fix the bulk mass parameters $c_i$ to a constant value and fit the Yukawa couplings $\lambda_i$; in the other, we fix $\lambda_i$ to a constant value and vary $c_i$. This strategy allows us to assess which combinations of bulk parameters are compatible with existing data. We also explored the fully general scenario in which both $c_i$ and $\lambda_i$ are fitted simultaneously. In this case, the model exhibited enough flexibility to always accommodate the data, and no regions of parameter space could be excluded.

We define the combined log-likelihood as a sum over the individual contributions of the considered experiments:
\begin{equation}
\log \mathcal{L}_{\text{tot}} = \sum_i \log \mathcal{L}_i,
\end{equation}
where each $\mathcal{L}_i$ corresponds to the (binned) likelihood for a given experiment, depending on both oscillation parameters and relevant systematics.

For each experimental dataset, we validated our simulation pipeline by reproducing the published standard oscillation results. We then scanned the parameter space of the extra-dimensional model, varying $R$ and $m_\mathrm{lightest}$ along with either the bulk mass parameters $c_i$ or the Yukawa couplings $\lambda_i$, while profiling over all nuisance parameters and minimizing the negative log-likelihood at each point. Exclusion contours are determined by computing the test statistic:
\begin{equation}
\Delta \chi^2 = -2 \log \left( \frac{\mathcal{L}{\text{BSM}}}{\mathcal{L}{\text{SM}}} \right),
\end{equation}
where $\mathcal{L}{\text{SM}}$ is the best-fit likelihood under the three-flavor model and $\mathcal{L}{\text{BSM}}$ the likelihood for the given beyond the SM hypothesis. We assume Wilks' theorem holds to convert $\Delta \chi^2$ values into confidence intervals.

This analysis allows us to place bounds on the compactification radius $R$ as a function of $m_\mathrm{lightest}$, given a combination of the parameters ($c_i$, $\lambda_i$). We run two separate analyses for the two possible neutrino mass orderings, NO and IO, respectively. The results of this data analysis are presented and discussed in the next section.

\section{RESULTS \label{sec: results}}

\begin{figure}
\centering
   \includegraphics[width=0.5\columnwidth]{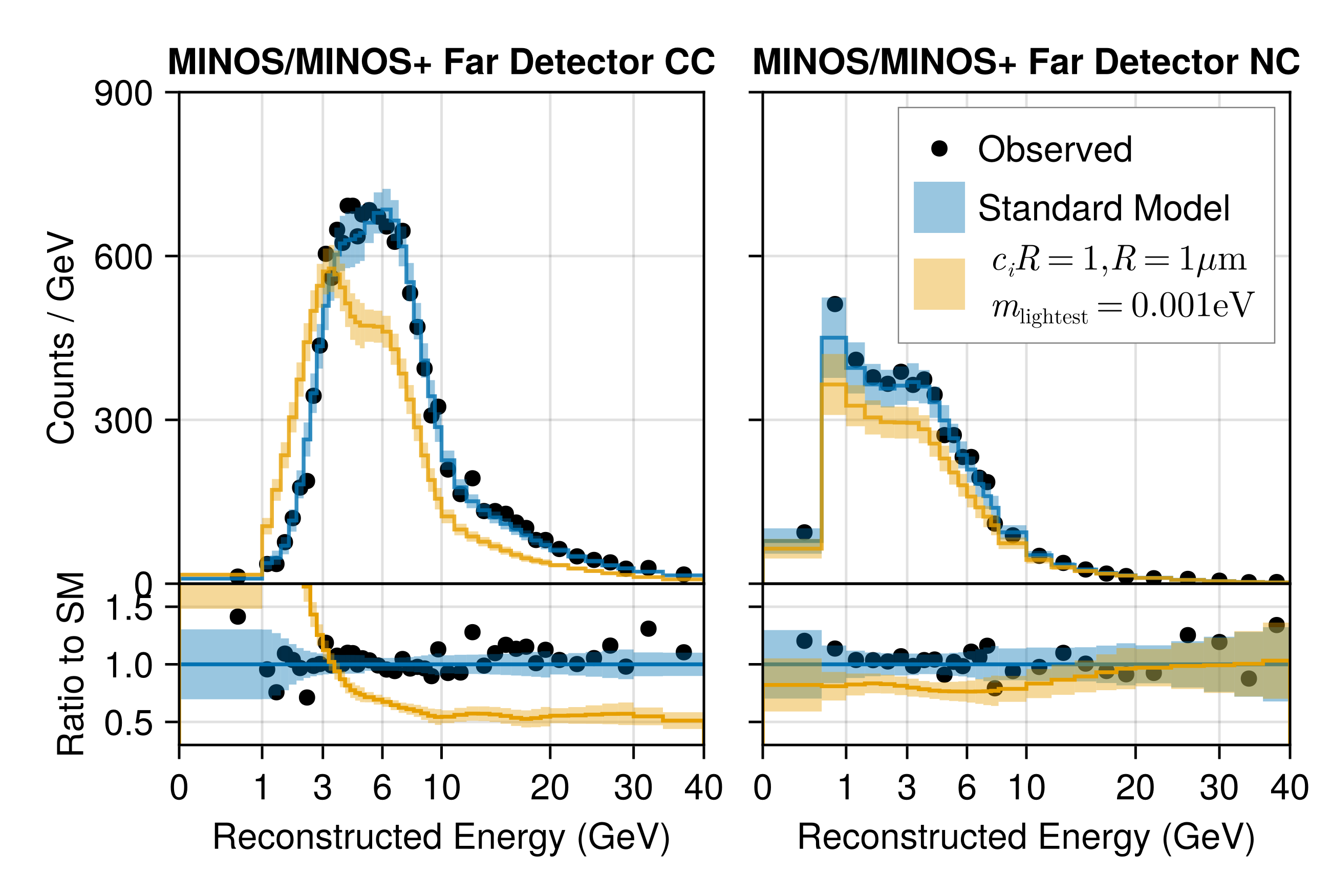}
   \includegraphics[width=0.5\columnwidth]{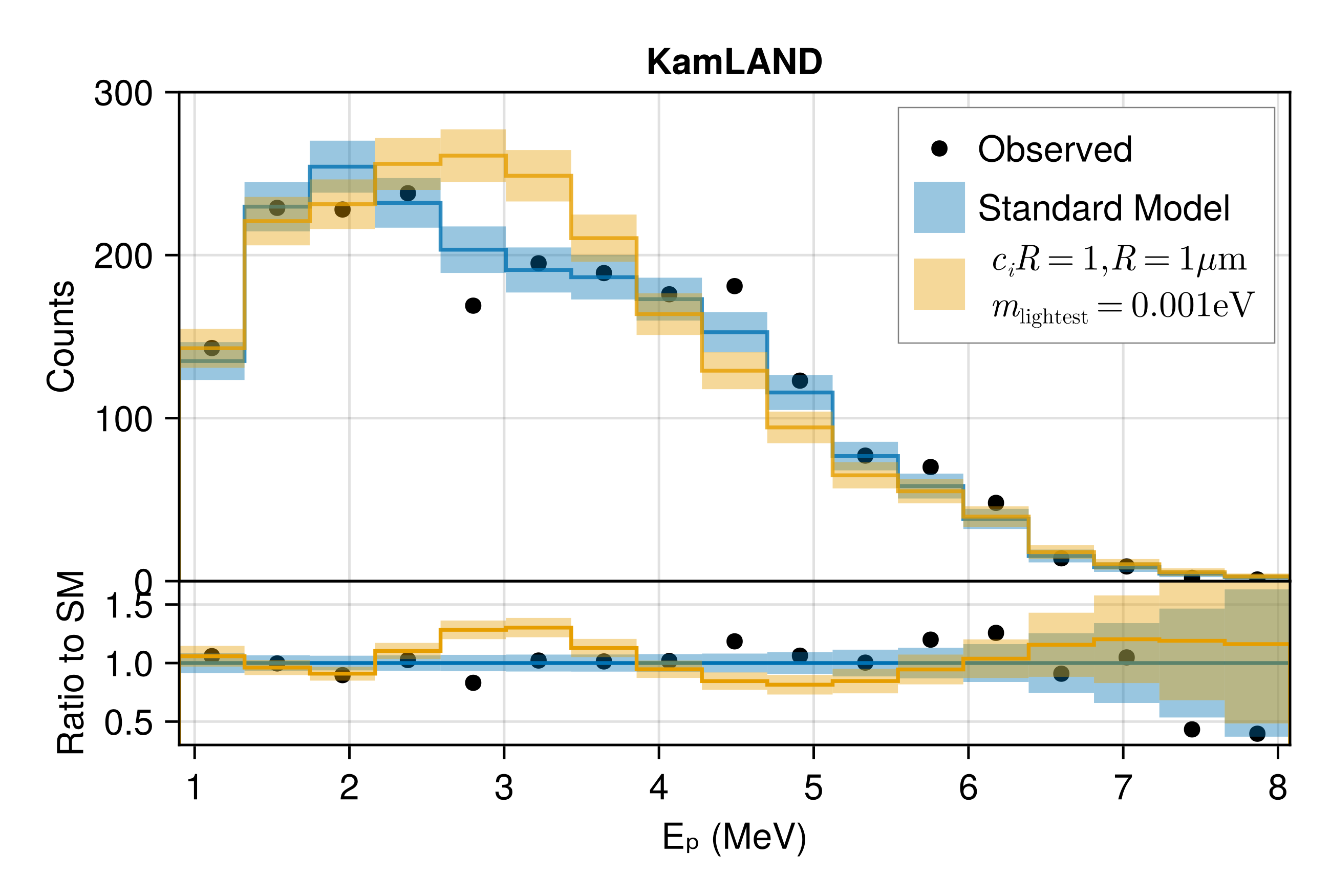}
   \includegraphics[width=0.5\columnwidth]{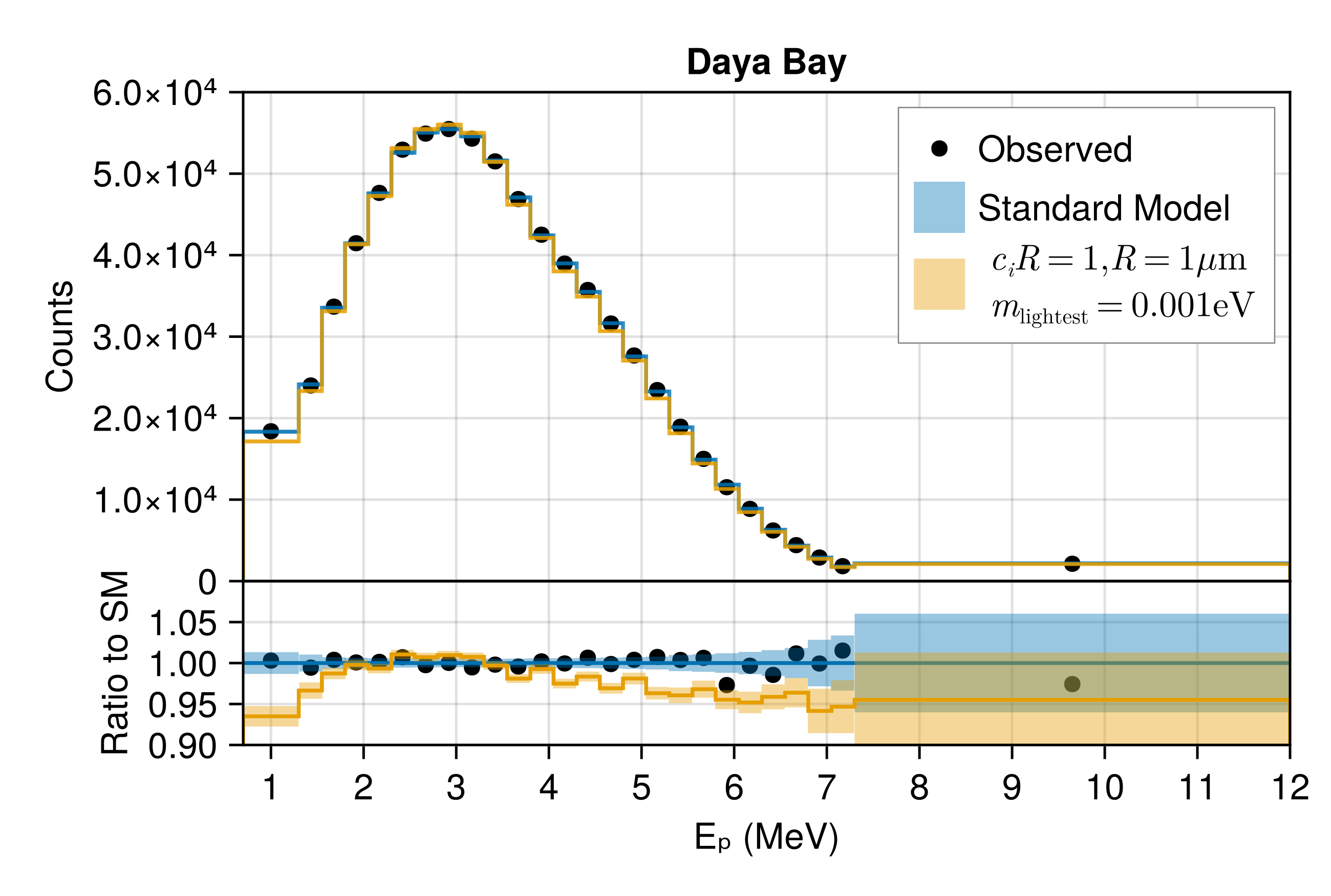}
  \caption{Comparison of the SM bestfit (blue) vs. the bestfit of an example point in our model parameter space ($R=1\,\mathrm{\mu m}, c_iR=1, m_\mathrm{lightest} = 0.001\,\mathrm{eV}$) assuming NO, for the MINOS/MINOS+ (top), KamLAND (middle), and Daya Bay (bottom) datasets. The solid lines represent the mean expectations, and the bands denote the 68\% statistical uncertainties.}
  \label{fig: DD Experiments}
\end{figure}

Before presenting the exclusion limits, we show the binned energy spectra that enter our fit. Figure~\ref{fig: DD Experiments} displays the observed event counts as a function of reconstructed neutrino energy for MINOS/MINOS+, KamLAND, and Daya Bay. In each case, we compare the data to the SM prediction and to a representative set of parameters ($R=1\,\mathrm{\mu m}, c_iR=1, m_\mathrm{lightest} = 0.001\,\mathrm{eV}$) and assuming NO.

These plots illustrate how the new physics affects the spectral shape and normalization, and demonstrate the level of agreement between theory and data that underlies the statistical results in this work.

\begin{figure*}[htb]
\centering
   \includegraphics[width=0.90\textwidth]{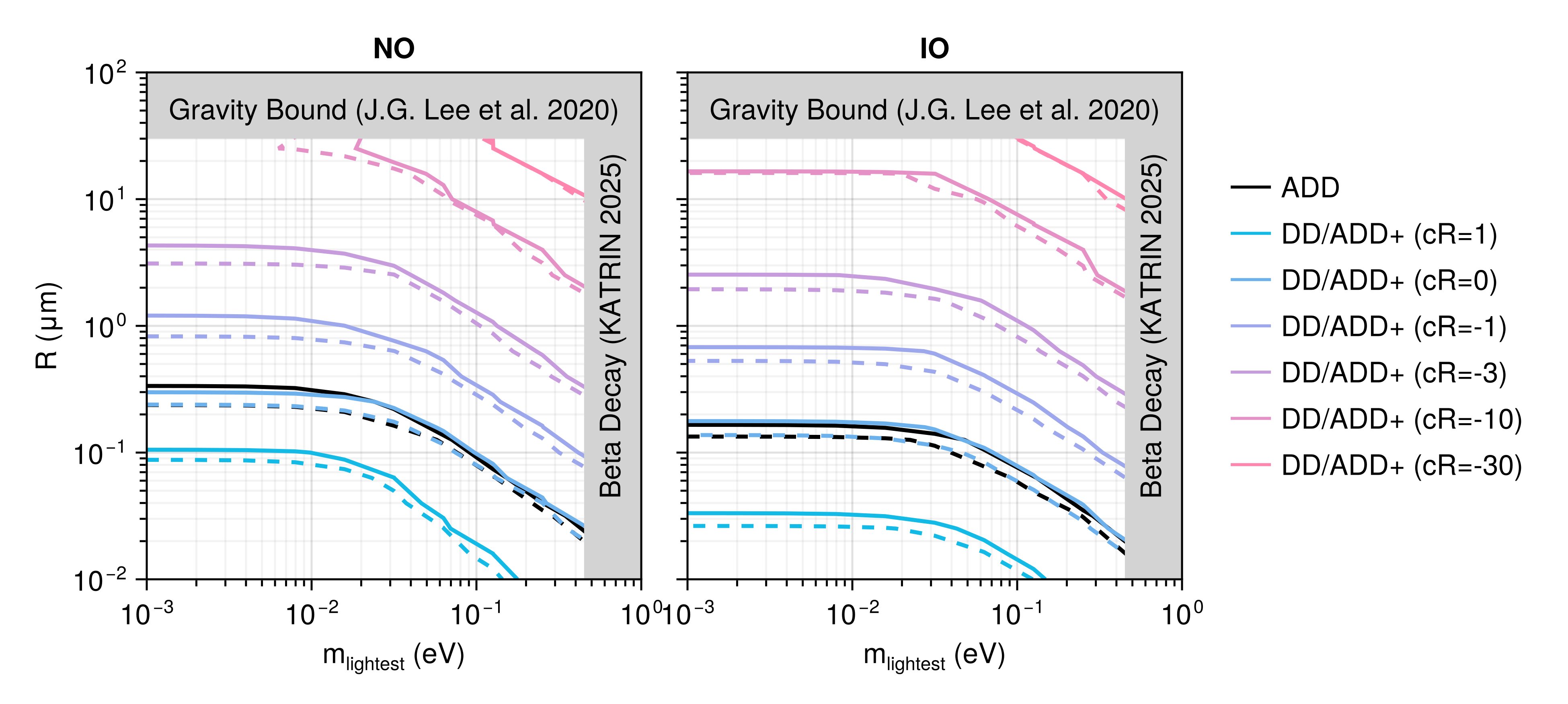}
    \caption{Exclusion limits at 90\% CL (dashed) and 99\% CL (solid) on the compactification radius $R$ as a function of the lightest neutrino mass $m_{\text{lightest}}$, shown for both normal ordering (left) and inverted ordering (right). Each curve corresponds to a fixed value of the dimensionless bulk mass parameter $cR$. The black line denotes the ADD benchmark case with $cR = 0$. The gray shaded regions indicate existing external constraints: the upper region is excluded by testing deviations from Newtonian gravity, while the vertical band reflects the upper bound to the effective electron neutrino mass. Together, these bounds define the edge of the physically viable parameter space.}
  \label{fig: R vs mlightest cR fixed}
\end{figure*}

\begin{figure*}[htb]
\centering
   \includegraphics[width=0.90\textwidth]{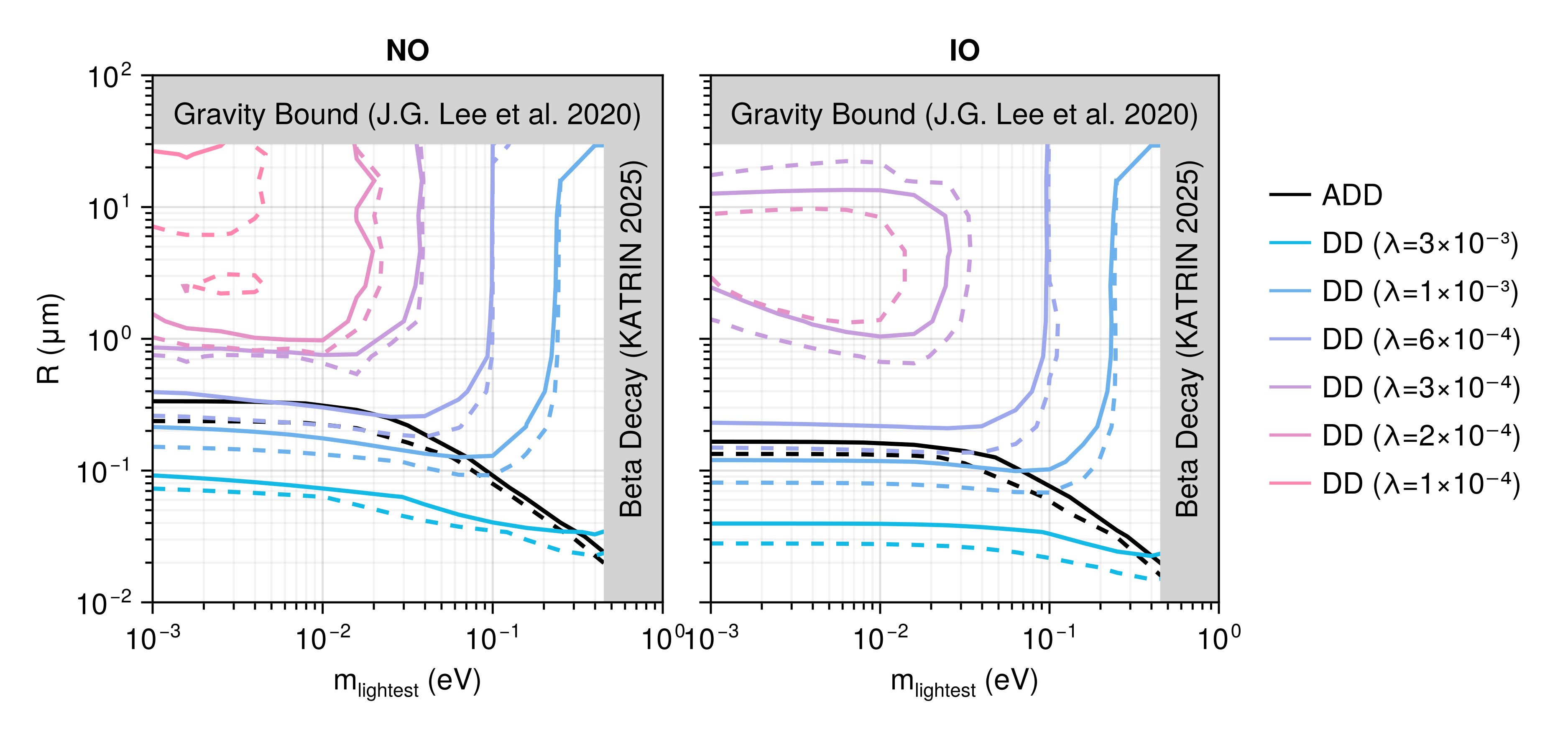}
   \includegraphics[width=0.90\textwidth]{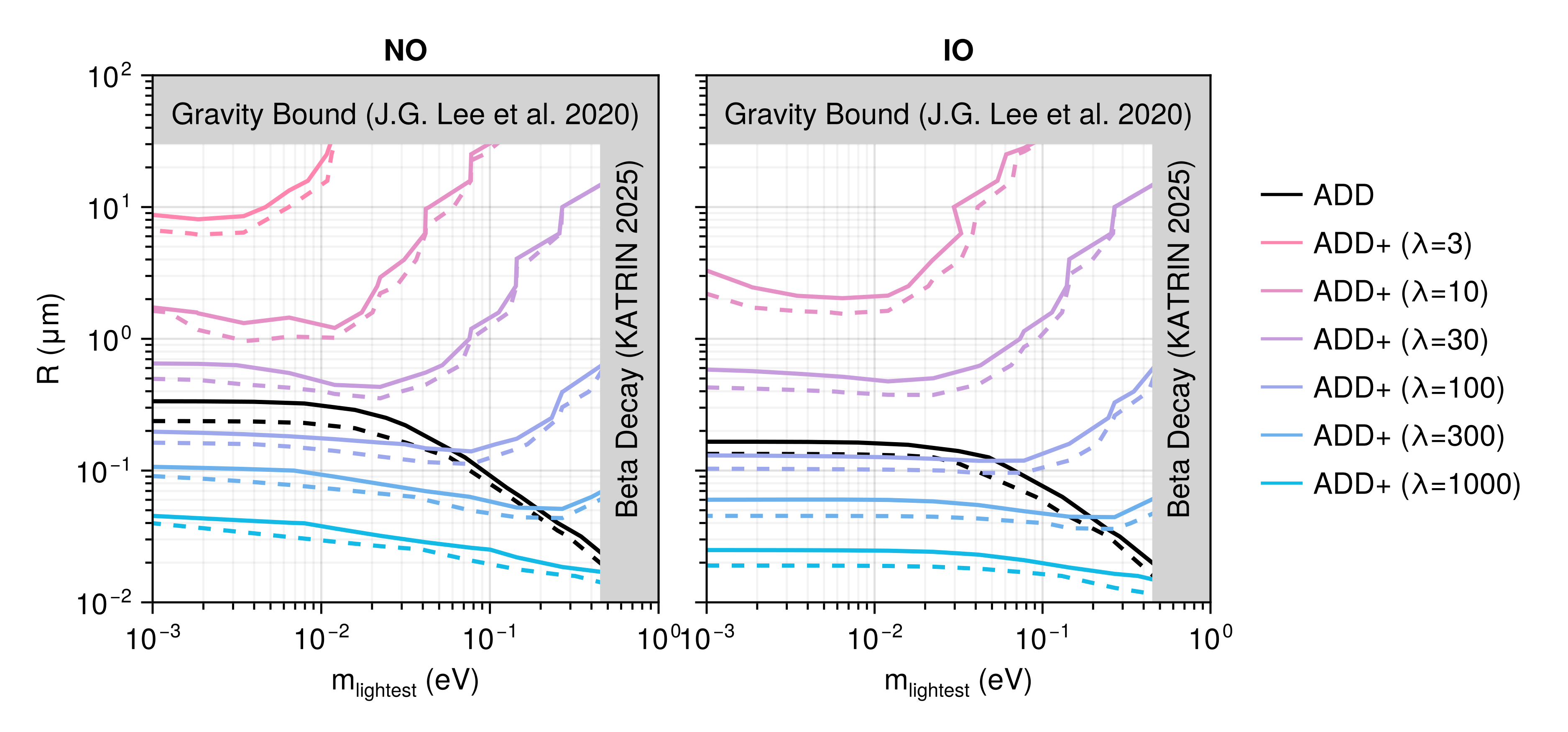}
    \caption{Exclusion limits at 90\% CL (dashed) and 99\% CL (solid) on the compactification radius $R$ as a function of $m_{\text{lightest}}$, for various fixed values of the Yukawa coupling $\lambda$. The top panel shows results for the DD scenario, while the bottom panel corresponds to ADD+. The black line again represents the ADD benchmark. External constraints from gravity tests and beta-decay experiments are indicated by the gray shaded regions.}
  \label{fig: R vs mlightest lambda fixed}
\end{figure*}

The exclusion limits obtained from our global fit are shown in Figures~\ref{fig: R vs mlightest cR fixed} and~\ref{fig: R vs mlightest lambda fixed}, for both normal ordering (NO) and inverted ordering (IO). These plots present upper bounds on the compactification radius $R$ as a function of the lightest neutrino mass $m_{\text{lightest}}$, under different model assumptions for the extra-dimensional parameters. In both figures, we also indicate external constraints from non-oscillation experiments: a vertical gray band shows the newest results of the KATRIN experiment to the effective electron neutrino mass, which roughly translates into values of $m_{\text{lightest}} \gtrsim 0.45$~eV at 90\% CL \cite{aker2406direct}, both for NO and IO. Meanwhile, the shaded horizontal region marks the bound coming from experiments that search for deviations in the $1/r^2$ law of the gravitational force, which currently constrain the compactification radius to $R \lesssim 30~\mu$m \cite{Lee:2020zjt}. These external limits serve to delineate the physically viable region of parameter space, independently of the oscillation framework.

In Figure~\ref{fig: R vs mlightest cR fixed}, we fix the dimensionless bulk mass parameters $c_iR$ and show results for various choices. Positive values of $c_iR$ yield much stronger constraints because the necessary dimensionless Yukawa couplings $\lambda_i$ to generate the correct SM mass splittings $\Delta m^2_{21}$ and $\Delta m^2_{31}$ grow exponentially with larger $c_iR$, as can be seen from equation (\ref{eq: Y0i}). Therefore, to match the observed mass splittings in current data, one is forced to introduce a large mixing of active neutrinos with the KK states, which is governed by the Yukawa couplings of equation (\ref{eq: Yni}). Negative values of $c_iR$ pose much weaker constraints on $R$, reflecting the decoupling of the KK tower due to suppressed mixing angles. Notably, we confirm that the ADD benchmark case ($c_iR = 0$) reproduces existing bounds reported in \cite{forero2022large}.

In Figure~\ref{fig: R vs mlightest lambda fixed}, we instead fix the Yukawa couplings $\lambda_i$ and probe different values, separately for ADD+ and DD. As expected, smaller values of $\lambda_i$ lead to weaker constraints on $R$, since smaller Yukawa couplings directly suppress the mixing between active and sterile KK modes. For large enough Yukawa couplings (e.g. $\lambda_i \gtrsim 10^{-3}$ in the case of DD), the derived bounds approach and in some regions surpass the existing ADD limits. In contrast, for $\lambda_i \lesssim 10^{-4}$, the mixing is too weak to leave a detectable imprint, and neutrino data fail to constrain the radius. Note that in the DD scenario, for smaller $\lambda_i$, see for instance $\lambda_i = 3 \cdot 10^{-4}$ for IO, the upper region of our parameter space reopens (up to the gravity bound) due to the scaling $M_f \sim R^{-1/3}$ from equation~(\ref{eq: Mf for DD}), which leads to a further reduction of the mixing angle between active and sterile states. The shape of the exclusion curves under the ADD+ model differs, as there is no such $R$ dependence of $M_f$, which is fixed to 10\,TeV.

In all diagrams, IO limits are generally stronger at low $m_\mathrm{lightest}$ due to greater BSM deviations from the SM, as shown in Figure~\ref{fig: oscillograms prob}. For $m_\mathrm{lightest} \gtrsim 0.1,\mathrm{eV}$, the difference between NO and IO becomes negligible.

\section{CONCLUSIONS \label{sec: conslusions}}

We have presented a global neutrino oscillation analysis of models with a single large extra dimension containing right-handed neutrinos with bulk Dirac masses. Our study covers two theoretically motivated realizations: the ADD+ framework and the Dark Dimension (DD) scenario. In both cases, the presence of a Kaluza–Klein tower modifies active–sterile mixing and can leave observable imprints in oscillation data.

Using data from MINOS/MINOS+, KamLAND, and Daya Bay, we derived bounds on the compactification radius $R$ by varying the oscillation parameters together with either the bulk masses or the brane-localized Yukawa couplings. Large positive bulk masses need large mixings between active and KK states to reproduce the observed mass splittings and  therefore tighten constraints, while large negative masses suppress mixing and relax them. Similarly, small Yukawa couplings weaken sensitivity, especially in the DD case where $M_f$ decreases with $R$.

Current oscillation data leave parts of the parameter space unconstrained, particularly for small $\lambda_i$ or strongly negative $c_iR$. These regions are further limited by independent bounds from KATRIN and short-range gravity tests, together defining the physically viable parameter space.

Beyond setting exclusion limits, our analysis maps out the regions in $(R, \lambda_i)$ and $(R, c_iR)$ space that remain viable given current constraints. This identification of allowed parameter space provides useful guidance for model building, indicating which combinations of compactification radius, Yukawa coupling, and bulk mass are still consistent with data. Future high-precision oscillation experiments, complemented by next-generation beta decay and gravity tests, will be able to further probe and potentially close these regions, offering a sharper picture of the role that large extra dimensions with bulk neutrinos might play in particle physics.

\section{Acknowledgments}
We thank Irene Valenzuela for encouraging us to publish our results. The work of M.E.~was supported by ANR grant ANR-23-CE31-0024 EUHiggs. The work of P.E. and A.Z. has been supported by the Deutsche Forschungsgemeinschaft (DFG, German Research Foundation) under Germany's Excellence Strategy – EXC-2094 – 390783311, and the SFB 1258 – 283604770
\clearpage

\bibliographystyle{unsrt}
\bibliography{mybib}
%


\end{document}